\documentclass[10pt,twocolumn]{article}

\usepackage[margin=0.72in,columnsep=0.24in]{geometry}
\usepackage[T1]{fontenc}
\usepackage[utf8]{inputenc}
\usepackage{microtype}
\usepackage{booktabs}
\usepackage{tabularx}
\usepackage{enumitem}
\usepackage{xcolor}
\usepackage{listings}
\usepackage{amsmath}
\usepackage{amssymb}
\usepackage{graphicx}
\usepackage{url}
\usepackage[hidelinks]{hyperref}

\definecolor{codegray}{rgb}{0.96,0.96,0.96}
\setlist{nosep,leftmargin=*}
\newcommand{\systema}{Workflow A}
\newcommand{\systemb}{Workflow B}
\newcommand{\systemc}{Workflow C}

\title{Trusted Workflow Relays:\\
Cross-Tenant Email Abuse and Composable Red Team Initial-Access Primitives in Multi-Tenant Clouds}
\author{Priyank Nigam\\
Microsoft Security Engineering}
\date{August 2026}
\begin{document}
\maketitle

\begin{abstract}
Cloud applications routinely send notifications through provider-operated mail
identities.  This delegation is convenient and usually improves deliverability,
but it also separates the actor who supplies notification parameters from the
service principal that originates the message.  We present a case study of this
\emph{trusted workflow relay} pattern in multi-tenant cloud services.  In three
responsibly disclosed and remediated cross-tenant notification workflows, an
authenticated actor could influence recipients across tenant boundaries and, to
different degrees, control message content that a trusted provider service
delivered.  In the first workflow, backend requests bypassed a user-interface
length restriction, raw HTML and CSS survived into the delivered message, HTTP(S)
links rendered, and CSS could visually suppress service-controlled text.
Tested iframes and non-web URI schemes
were rejected.  A second workflow combined missing recipient-tenant validation
with attacker-controlled subject and HTML fields.  A third workflow, an approval
application, combined weak access control, sequential object identifiers,
missing action authorization, and incomplete token validation, and shows how
notification abuse composes with authorization failures.

The pattern is analogous to a classical unauthenticated SMTP open relay, but the
failure has moved up the stack: the initiating actor is authenticated and the
cloud provider is the legitimate sender, yet application-layer authorization still
fails to constrain who may cause it to send what to whom.  We formally define a
trusted workflow relay as a delivered, service-authentic message for which the
application-level send-authorization predicate is false.  We then give a test
matrix for notification pipelines, map the primitive to MITRE ATT\&CK techniques
for attachment-free phishing, and connect it to device-authorization
phishing documented in RFC~8628 and recent threat campaigns.
The results show why the standard email authentication mechanisms---Sender Policy
Framework (SPF), DomainKeys Identified Mail (DKIM), and Domain-based Message
Authentication, Reporting, and Conformance (DMARC)---can correctly authenticate a
message yet cannot establish that an application-level send was authorized.  We conclude with
controls for tenant binding, typed templates, object-level authorization, token
audience validation, and identity telemetry.
\end{abstract}

\section{Introduction}

Traditional E-mail relays operate at the transport layer.  An improperly configured
relay may accept mail from an unauthorized party and forward it to an arbitrary
recipient.  Cloud software has shifted much business email origination above that
layer.  A user clicks \emph{share}, \emph{invite}, \emph{approve}, or \emph{notify};
an application backend then constructs a message and a provider-controlled service
identity sends it.  The resulting message is not forged.  Its infrastructure and
domain authentication can be entirely legitimate.

This change creates a different authorization problem.  A multi-tenant application
must bind four elements before sending: the initiating principal, its tenant, the
target object and recipient, and the content permitted in the notification.  A
failure at any of these boundaries can turn a legitimate notification service into
an application-layer relay.  Existing email authentication mechanisms do not make
this decision.  The Sender Policy Framework (SPF) authorizes the hosts that may
send mail for a domain, DomainKeys Identified Mail (DKIM) attaches a cryptographic
signature that associates a signing domain with message content, and Domain-based
Message Authentication, Reporting, and Conformance (DMARC) evaluates whether those
identifiers align with the author domain and applies the domain owner's policy
\cite{rfc7208,rfc6376,rfc9989}.  When the provider itself originates the message,
those checks can pass while the business operation that caused the send was
unauthorized.

This paper reports findings from security assessments of multi-tenant cloud
workflows and places them in a common model.  The specific vulnerabilities were
responsibly disclosed and remediated.  Product details and identifiers that are
not needed to understand the failure mode are generalized.  Our contributions are:

\begin{itemize}
  \item a formal definition of the \emph{trusted workflow relay} as the
  conjunction of service-authentic delivery and failed application-level
  authorization, together with a decomposition of the required authorization
  predicates;
  \item a qualitative case study of three cross-tenant notification workflows,
  including a content-rendering test matrix derived from delivered messages;
  \item an analysis of how object-level authorization and token-validation flaws
  in the third workflow compose with trusted notification channels;
  \item a mapping of the primitive to MITRE ATT\&CK for attachment-free phishing
  that identifies both the techniques it never needs and the recommended
  mitigation that fails by construction; and
  \item a set of engineering and detection controls spanning notification
  origination, content rendering, authorization, and identity telemetry.
\end{itemize}

\section{Background and Threat Model}

\subsection{Formal definition: from SMTP relays to workflow relays}

A classical SMTP open relay accepts and forwards messages without an appropriate
authorization decision.  The workflows studied here did not expose a general SMTP
interface.  Instead, an authenticated user invoked a business operation, and a
backend service generated mail through a privileged provider identity.  We use
\emph{relay} to describe the security effect, not protocol equivalence.

Let $u$ be an initiating principal in tenant $T_u$, $r$ a recipient in tenant
$T_r$, $o$ a shared or actioned object, $c$ attacker-influenced content, and $s$ a
trusted sending service.  Let $q=(u,T_u,o,r,T_r,c)$ denote a notification request.
The application's authorization decision for that request is

\begin{equation}
\label{eq:authorization}
\begin{split}
\mathsf{Authorized}_s(q) \equiv{}& \mathsf{MayAct}(u,T_u,o) \\
  &\land\ \mathsf{MayTarget}(u,T_u,o,r,T_r) \\
  &\land\ \mathsf{MayOriginate}_s(u,T_u,o,r,T_r) \\
  &\land\ \mathsf{SafeContent}_s(c).
\end{split}
\end{equation}

Here, the predicates respectively authorize the actor's operation on the object,
the recipient relationship, the privileged notification side effect, and the
content admitted to the service-owned template.

\paragraph{Definition (trusted workflow relay).}
A service $s$ exposes a trusted workflow relay if there exists a request $q$ and
message $m$ such that

\begin{equation}
\label{eq:relay}
\boxed{\begin{gathered}
\exists\,q,m:\ \mathsf{Delivered}_s(q,m) \\
{}\land\ \mathsf{Authentic}_s(m)
\land\ \neg\mathsf{Authorized}_s(q)
\end{gathered}}
\end{equation}

$\mathsf{Authentic}_s(m)$ means that the message genuinely originates from and
authenticates as the provider service; it does not assert that the initiating
application action was authorized.  Thus, ``trusted'' describes the message's
service identity, while ``relay'' describes the unauthorized use of that identity's
delivery authority.  Equation~\ref{eq:relay} is satisfied when any predicate in
Equation~\ref{eq:authorization} fails.  The most consequential cases combine a
recipient-boundary failure with a content-boundary failure.

\subsection{Attacker capabilities}

Our primary attacker has a valid low-privilege account in a tenant they control or
an otherwise ordinary consumer (non-enterprise) account accepted by the target service.  The
attacker can use browser developer tools or an intercepting proxy to replay and
modify requests that their account is already able to issue.  The attacker does
not possess a provider signing key, compromise a mail server, or control the
recipient tenant.  For the approval-workflow case, we additionally consider an
attacker able to alter a numeric object identifier in a URL.

We assume the recipient's mail system may use SPF, DKIM, DMARC, reputation, and
content analysis.  These controls are relevant but cannot replace authorization at
the workflow that originates the message.  Prior work has shown that provider
handling and user-interface cues materially affect spoofing outcomes
\cite{hu2018spoofing}; our work examines the distinct case in which the sender is
authentic but the send authority is over-broad.

\section{Methodology}

\subsection{Assessment procedure}

We assessed the workflows as black-box application surfaces.  For each workflow,
we first exercised its intended user interface and captured the corresponding HTTP
request.  We then changed one security-relevant field at a time and observed (1)
the HTTP response, (2) whether a message was delivered to a researcher-controlled
mailbox, and (3) how the resulting body rendered in the tested mail client.  The
input classes were recipient tenant, directory object identifier, subject, body
length, markup, style, embedded frame, and link scheme.

Tests used accounts, tenants, objects, and mailboxes controlled by or explicitly
available to the researcher.  We used benign proof text and links.  We did not send
unsolicited messages, collect credentials, or use delivered messages for a live
phishing campaign.  In the approval workflow, a missing authorization check on a
state-changing action was identified but not exercised against a real pending
record; we therefore report that consequence as potential impact, not an observed
state change.

\subsection{Evidence classification}

We label evidence in three categories.  \emph{Observed} means a controlled request
or delivered message demonstrated the behavior.  \emph{Inferred} means a reachable
code or UI path supported an impact that was deliberately not exercised.
\emph{Literature-derived} means the claim comes from standards, vendor guidance, or
threat-intelligence reporting.  This distinction is important because successful
delivery does not itself establish inbox placement across providers, user
interaction, or account compromise.

\subsection{Reproducibility constraints}

The services were remediated after disclosure, so the original behavior is not
expected to reproduce on current production deployments.  To avoid exposing tenant
data or stale exploit details, this paper includes sanitized request shapes and the
test matrix rather than raw captures.  Exact product versions were not available
from the user-facing services.  These constraints favor a qualitative security
case study over a prevalence measurement.

\section{Case Study: Notification Content and Client Rendering}

\subsection{Workflow and boundary failure}

\systema{} was a multi-tenant application-sharing workflow.  The normal UI allowed
a user to choose a principal, assign a view role, and optionally add a short custom
message.  A provider-operated backend generated the notification; the user did not
connect to SMTP or choose the envelope sender.  The relevant request shape was:

\begin{lstlisting}
{"put":[{"properties":{
  "NotifyShareTargetOption":"Notify",
  "roleName":"CanView",
  "principal":{"id":"<recipient-object-id>",
               "type":"User"}}}],
 "emailCustomizations":{"message":"<content>"}}
\end{lstlisting}

The UI attempted to constrain message length and rejected obvious HTML characters.
Those checks did not define the backend's effective policy.  A direct request with
more than 200 characters received a success response, and raw markup supplied to
the backend appeared in delivered messages.

\subsection{Rendering matrix}

Table~\ref{tab:rendering} summarizes the controlled tests represented in the
research artifacts.  ``Rendered'' means the supplied feature had the intended
visual effect in the tested delivered message.  A failed test narrows the observed
surface; it does not prove that all variants of that class were blocked.

\begin{table*}[t]
\centering
\caption{Observed behavior in \systema{}.  Results are qualitative and specific to
the tested workflow and client at assessment time.}
\label{tab:rendering}
\begin{tabularx}{\textwidth}{@{}p{0.25\textwidth}p{0.13\textwidth}X@{}}
\toprule
Test & Result & Security interpretation \\
\midrule
Body longer than UI limit & Accepted & Client-side length policy was not enforced by the backend. \\
Plain text & Rendered & Baseline custom-message behavior. \\
Encoded markup & Not rendered as markup & The tested encoded representation did not become active HTML. \\
Raw HTML tags & Rendered & Message data reached an HTML-capable sink. \\
Arbitrary iframe & Blocked & The tested frame did not render. \\
HTTP(S) link & Rendered & Attacker-selected web destination was clickable. \\
Non-web URI schemes & Blocked & Tested file, FTP, SSH, and other schemes did not render as usable links. \\
CSS rules & Applied & Attacker-controlled style affected presentation. \\
CSS positioning & Applied & Content could be repositioned relative to trusted template elements. \\
Iframe with tested allow-listed host & Blocked & No frame execution was observed in that variant. \\
Hide service-controlled text & Applied & CSS could suppress or visually truncate trusted explanatory text. \\
Combined HTML, CSS, and HTTPS call to action & Rendered & A provider-originated message could present attacker-selected copy and a prominent external link. \\
\bottomrule
\end{tabularx}
\end{table*}

The key result was not script execution.  Email clients commonly disable
JavaScript, yet HTML and CSS remain security-relevant presentation languages.
Prior client research has likewise shown that HTML/CSS interpretation can create
security consequences without JavaScript \cite{poddebniak2018efail}.  Here, CSS
could hide service-controlled context and place attacker-controlled content in its
visual position.  This creates an integrity failure in the message template even
when active content such as an iframe is blocked.

\subsection{Impact and remediation}

The provider, not the initiating user, authenticated the resulting mail.  Thus,
anti-spoofing success was an expected property of the message rather than evidence
that its custom content had been authorized.  We observed delivery to a
researcher-controlled mailbox and a recognizable provider sender presentation.  We
did not conduct a controlled comparison of spam verdicts or inbox rates and make no
quantitative deliverability claim.

The durable fix is to enforce policy at the server-side composition boundary:
bind recipients to authorized tenant relationships; treat custom text as data;
escape it for the output context; and expose a small typed template vocabulary
instead of arbitrary HTML or CSS.  A client-side character check or length limit is
not a security boundary.

\section{Case Study: Cross-Tenant Trusted-Sender Abuse}

\systemb{} was a second Microsoft-hosted, multi-tenant collaboration workflow.  A
share operation accepted both a directory object identifier and an email address,
then sent a notification from a provider system domain.  The assessed backend did
not verify that the recipient resolved inside the sharer's tenant or another
explicitly authorized collaboration relationship.  It also accepted a custom
subject and HTML message body.  The request can be represented as:

\begin{lstlisting}
POST /api/project/share
{
  "AADObjectId": "<supplied-guid>",
  "Email": "recipient@other-tenant.example",
  "CustomSubject": "<supplied-subject>",
  "CustomMessage": "<supplied-html>"
}
\end{lstlisting}

The recipient-boundary and content-boundary failures composed.  An actor in one
tenant could cause the provider workflow to originate a customized message to a
mailbox in another tenant.  In contrast to lookalike-domain or From-header
spoofing, the provider system was the actual originator.  SPF, DKIM, and DMARC are
therefore not expected to reject the message merely because a tenant user abused
the application that requested it.  This is consistent with the standards' scope:
domain authentication does not validate the legitimacy of message content or the
application-level decision to send it \cite{rfc7208,rfc6376,rfc9989}.

The remediation validated both the email address and directory object identifier
against the host tenant before sending and rejected recipients outside the
authorized boundary.  Both values must be resolved and compared server-side;
checking only a display string or trusting a caller-supplied tenant identifier
leaves room for inconsistent bindings.

\section{Case Study: Authorization Composition in an Approval Workflow}

\systemc{} was a third cross-tenant notification workflow.  It managed invoice or
access-approval records, and its approve and reject actions caused a provider
identity to notify requesters and approvers who could sit outside the acting
account's tenant.  Its administration page was
intended for employees and a limited set of onboarded suppliers, but an ordinary
consumer (non-enterprise)account could reach the page.  Records were selected by a sequential
\texttt{requestId}:

\begin{lstlisting}
GET /Administration/AccessRequestDetails.aspx
    ?type=approval&requestId=<N>
\end{lstlisting}

Changing $N$ exposed records outside the account's authorization scope.  This is an
insecure direct object reference: user-controlled input selected an object without
a corresponding object-level authorization decision \cite{owaspIDOR}.  The page
also exposed approve and reject actions without an adequate role check.  We did not
exercise those actions on live records.  Because such actions generated trusted
workflow notifications, unauthorized mutation could also cause provider-originated
mail with business consequences.

The same assessment surfaced an incomplete access-token validation pattern.  A
multi-tenant API cannot treat a valid signature or tenant claim as sufficient.  A
resource server must verify that the token's \texttt{aud} claim identifies that
resource; otherwise a token minted for another application may be accepted by a
confused deputy \cite{msAccessTokens,msClaimsValidation}.  It must then authorize
the subject and calling application for the selected tenant, object, and action.

This case illustrates composition.  A valid identity is not necessarily an
authorized employee; an authenticated user is not necessarily allowed to read the
selected object; permission to read one object does not imply permission to change
it; and permission to change an object does not automatically justify sending a
notification.  Authorization should be deny-by-default and evaluated on every
request \cite{owaspAuthorization}.

\section{Formal Definition and Boundary Model}

Equation~\ref{eq:relay} separates the property established by sender
authentication, $\mathsf{Authentic}_s(m)$, from the application authorization
property that must hold before delivery, $\mathsf{Authorized}_s(q)$.  Across the
cases, that authorization property crosses several independently managed
boundaries.  Table~\ref{tab:boundaries} operationalizes the formal definition as a
review taxonomy.

\begin{table*}[t]
\centering
\caption{Boundary-oriented review model for cloud notification workflows.}
\label{tab:boundaries}
\begin{tabularx}{\textwidth}{@{}p{0.18\textwidth}p{0.27\textwidth}X@{}}
\toprule
Boundary & Failure pattern & Required invariant \\
\midrule
Initiator & Any authenticated or consumer (non-enterprise)identity is treated as an approved actor. & Authorize immutable subject and tenant identifiers for the operation. \\
Object & Caller-supplied or sequential IDs select records outside the actor's scope. & Enforce object-level authorization after object resolution, for reads and writes. \\
Recipient & Email and directory identifiers are accepted without tenant binding. & Resolve both values server-side and verify an allowed same-tenant or explicit cross-tenant relationship. \\
Content & User data reaches subject or HTML/CSS sinks with only UI validation. & Use contextual encoding and typed templates; prohibit arbitrary style and markup. \\
Origination & A privileged service identity signs mail caused by a weakly authorized action. & Treat sending as a privileged side effect with its own policy and audit event. \\
Token & Signature or issuer is checked but audience, tenant, subject, or actor is not. & Use maintained middleware and validate audience plus contextual authorization claims. \\
Telemetry & Logs record only the service sender, obscuring the initiating tenant user. & Correlate initiator, tenant, object, recipient tenant, template, and message identifier. \\
\bottomrule
\end{tabularx}
\end{table*}

The model also explains why individually reasonable controls can fail in
composition.  A sanitizer may block iframes but still permit CSS that hides trusted
text.  A mail gateway may correctly trust the sending domain but lack the tenant
context needed to judge the send.  An API may validate a token cryptographically
but accept the wrong audience.  Effective review therefore follows data and
authority from the initiating request through object access, template composition,
mail origination, and client rendering.

\section{Composition with Device-Code Phishing}
\label{sec:devicecode}

Device authorization is not a vulnerability in the assessed notification
workflows.  It is a standardized OAuth flow for devices that lack a suitable
browser or input mechanism \cite{rfc8628}.  We discuss it because public campaigns
show how a trusted delivery channel can compose with a legitimate identity flow to
create initial access.

In the legitimate flow, a client obtains a \texttt{device\_code}, a short
\texttt{user\_code}, and a verification URI.  A user authenticates on a second
device and enters the user code while the original client polls the token endpoint.
The client holding the device code receives the resulting token.  RFC~8628
explicitly identifies remote phishing: an attacker can initiate the flow and
convince a target to enter the attacker's code at the genuine verification page
\cite{rfc8628}.

The resulting composition is:

\begin{enumerate}
  \item the attacker initiates a device authorization request;
  \item a trusted or convincing notification carries the legitimate verification
  URI and attacker-controlled user code;
  \item the victim authenticates and satisfies MFA on provider infrastructure;
  \item the attacker, who retains the device code, polls the token endpoint and
  receives tokens for the approved client and scope.
\end{enumerate}

Microsoft and Volexity documented 2024--2025 campaigns using this pattern,
including real-time rapport building, Microsoft Teams and external-tenant themes,
and automatically refreshed device codes \cite{microsoftStorm2372,volexity2025}.
Microsoft further reported use of the Microsoft Authentication Broker client ID to
obtain a token for device registration, register an actor-controlled device, and
obtain a Primary Refresh Token \cite{microsoftStorm2372}.  These are
literature-derived observations; our notification testing did not request victim
tokens or reproduce campaign post-compromise behavior.

\section{Mapping to MITRE ATT\&CK}

The trusted workflow relay is not a new adversary technique.  It is a new way to
satisfy the preconditions of existing ones, and it does so without an attachment.
Table~\ref{tab:attack} maps the primitive and its device-code composition onto
ATT\&CK for Enterprise \cite{mitreAttack}.  The mapping is analytic: it describes
how the assessed defects would be used, and only the token-related rows are
supported by public campaign reporting
\cite{microsoftStorm2372,volexity2025}.

\begin{table*}[t]
\centering
\caption{Attachment-free phishing mapped to MITRE ATT\&CK for Enterprise.  Every
delivered message in the case studies carried text, markup, and links only.}
\label{tab:attack}
\begin{tabularx}{\textwidth}{@{}p{0.15\textwidth}p{0.30\textwidth}X@{}}
\toprule
Tactic & Technique (ID) & Realization through a trusted workflow relay \\
\midrule
Resource Development & Establish Accounts: Cloud Accounts (T1585.003) & The actor registers an ordinary tenant or consumer (non-enterprise) account that the multi-tenant service accepts; no mail server, domain, or signing key is acquired. \\
Resource Development & Acquire Infrastructure: Web Services (T1583.006) & The provider's own notification pipeline replaces attacker-controlled sending infrastructure, so no lookalike domain (T1583.001) is registered or aged. \\
Initial Access & Phishing: Spearphishing Link (T1566.002) & Attacker-supplied HTTP(S) anchors survived into the delivered body (Table~\ref{tab:rendering}).  ATT\&CK places consent phishing and device-code phishing under this sub-technique \cite{mitreT1566link}. \\
Initial Access & Phishing: Spearphishing via Service (T1566.003) & ATT\&CK already records abuse of the notification features of legitimate file-sharing services \cite{mitreT1566service}.  The relay generalizes that behavior to first-party cloud workflows, but the message arrives through the enterprise mail channel rather than around it. \\
Initial Access & Trusted Relationship (T1199) & The abused trust is the recipient's trust in the provider's sending identity rather than delegated administrative access \cite{mitreT1199}; adjacent to, not identical with, ATT\&CK's partner and service-provider framing. \\
Defense Evasion & Impersonation (T1656) & Attacker-controlled subject text and CSS that visually suppresses service-controlled text let the message read as a first-party notice. \\
Execution & User Execution: Malicious Link (T1204.001) & The victim clicks inside a message that both the mail gateway and prior user training treat as authentic provider correspondence. \\
Credential Access & Steal Application Access Token (T1528) & A device-authorization or consent link converts one click into tokens without a credential prompt \cite{mitreT1528}. \\
Persistence, Lateral Movement & Valid Accounts: Cloud Accounts (T1078.004); Application Access Token (T1550.001) & Follow-on access uses issued tokens and registered devices rather than malware, so endpoint-centric detection has no artifact. \\
\bottomrule
\end{tabularx}
\end{table*}

\paragraph{Techniques the primitive never needs.}
Spearphishing Attachment (T1566.001) and User Execution: Malicious File
(T1204.002) do not appear.  There is no file to detonate, so attachment sandboxing
and the Antivirus/Antimalware mitigation (M1049) have no object to evaluate, and
detection analytics keyed to attachment metadata or post-download process
lineage never fire.  Nothing needs to execute on the endpoint for the primitive to
succeed; the payload is a link and a pretext.

\paragraph{Where the recommended mitigation is a no-op.}
ATT\&CK's Software Configuration mitigation (M1054) for T1566.002 recommends SPF,
DKIM, and DMARC to filter messages by sender-domain validity
\cite{mitreT1566link}.  Equation~\ref{eq:relay} explains why that mitigation
cannot help here: $\mathsf{Authentic}_s(m)$ holds by construction, so every
sender-validity check the mitigation prescribes returns a pass.  The remaining
mapped mitigations---User Training (M1017), Restrict Web-Based Content (M1021),
and Audit (M1047)---are receiver-side and partial.  The control that would
actually falsify the technique, provider-side send authorization, has no ATT\&CK
mitigation, because ATT\&CK models adversary behavior against a defending
enterprise rather than authorization defects inside the sending service.

\paragraph{Detection implications.}
Because the attachment and endpoint stages are absent, detection weight shifts to
three signals: link analysis on messages that pass domain authentication,
sender-provenance telemetry that identifies the initiating tenant behind a service
identity, and identity telemetry for the token follow-on described in
Section~\ref{sec:devicecode}.  Only the provider can emit the second signal, which
is why the telemetry row of Table~\ref{tab:boundaries} is a detection prerequisite
and not merely an engineering hygiene item.

\section{Defensive Recommendations}

\subsection{Notification pipeline}

\begin{itemize}
  \item Model notification origination as a privileged action.  Check initiator,
  tenant, object, recipient relationship, and template policy immediately before
  enqueueing a message.
  \item Resolve recipient email and immutable directory identifiers server-side.
  Reject inconsistent pairs and cross-tenant recipients unless a documented
  collaboration policy explicitly permits them.
  \item Replace arbitrary subject and HTML fields with typed templates.  Encode
  text for its output context, allow only server-owned links or validated
  destinations, and do not allow tenant input to define CSS.
  \item Apply server-side length and schema constraints.  Exercise the backend
  directly in tests; UI restrictions are usability controls, not authorization.
  \item Record the causal chain: initiating principal and tenant, object, recipient
  and tenant, template identifier, message identifier, and authorization decision.
\end{itemize}

\subsection{Application and identity controls}

Every object read and state change requires an authorization decision after the
server resolves the target object.  Random identifiers reduce opportunistic
enumeration but do not replace access control.  For access tokens, use maintained
identity middleware and validate signature, issuer, lifetime, audience, tenant,
subject, calling application, and required scope or role as applicable
\cite{msAccessTokens,msClaimsValidation}.

For organizations that do not need device authorization, Microsoft recommends
blocking it through Conditional Access.  Where it is required, scope it narrowly
and first use report-only policy or sign-in logs to establish legitimate usage
\cite{msAuthFlows}.  Detection should baseline and correlate
\texttt{authenticationProtocol=deviceCode},
\texttt{originalTransferMethod=deviceCodeFlow}, device registration, anomalous
token use, and access to device-login URLs \cite{volexity2025,msAuthFlows}.  A
suspected compromise requires token/session revocation and investigation of newly
registered devices; revocation can take several minutes to propagate
\cite{msRevokeSessions}.

\subsection{Mail receiver controls}

Receivers should retain SPF, DKIM, and DMARC because they address important forms
of spoofing.  They should not treat authentication success as proof that arbitrary
content authorized by a multi-tenant application is benign.  Provider and SaaS
notification streams benefit from behavior-based controls such as unusual
recipient-tenant fan-out, novel external links, content-template divergence, and
reputation at a finer granularity than the parent sending domain.  Providers are in
the best position to expose an authenticated initiating-tenant signal to these
controls.

\section{Responsible Disclosure and Ethics}

The case-study vulnerabilities were reported through responsible disclosure and
were remediated before this preprint.  Testing was limited to benign content and
researcher-controlled resources.  No production records were modified to prove
the inferred approval impact, and no credentials or tokens were solicited.  The
paper omits live tenant identifiers, hostnames, and record values that are not
necessary to reproduce the reasoning.  Sanitization aims to preserve the
engineering lesson while reducing risk to historical or analogous deployments.

\section{Limitations}

This is a small, purposive case study, not a survey of cloud-service prevalence.
The tested services, mail client, and production versions may have changed, and
remediation prevents current replication.  The deck artifacts did not preserve a
complete set of raw Authentication-Results headers, spam verdicts, repeated trials,
or cross-provider delivery measurements.  Consequently, we report observed
delivery and rendering only and do not estimate inbox rate or click-through.

Some system details remain anonymized, which limits independent reproduction.
The content tests establish the behavior of the tested payloads, not the complete
HTML/CSS or URI-scheme policy.  The approval action's missing authorization was not
exercised on a live pending record.  The ATT\&CK mapping is an analytic
correspondence between the assessed defects and published technique definitions;
we did not observe an adversary executing these techniques through the assessed
workflows, and technique identifiers may be renumbered in later ATT\&CK releases.
Finally, the device-code section synthesizes
standards and public incident reporting rather than presenting a new campaign or
user study.

\section{Conclusion}

Cloud notification systems move email authority from end users to privileged
service identities.  That design can improve security and reliability, but only if
the application strongly binds the initiating principal, tenant, object, recipient,
and permitted content.  The assessed workflows broke those bindings in different
ways: cross-tenant recipients were insufficiently validated, rich content reached
mail-client rendering sinks, object identifiers were not authorization boundaries,
and token validation omitted essential context.

The central defensive lesson is that message authenticity and send authorization
are separate properties.  SPF, DKIM, and DMARC may all work as designed while a
trusted application sends an attacker-influenced message.  Providers should make
the application-level authorization decision explicit, test it at the backend,
carry its provenance into telemetry, and constrain notification content with typed
templates.  Defenders should then correlate that provenance with identity flows,
especially legitimate but phishable mechanisms such as OAuth device authorization.

\section*{Acknowledgments and Disclaimer}

The author thanks the engineering teams that investigated and remediated the
reported issues.  The views expressed are the author's own and do not necessarily
represent the views of Microsoft.  Product and company names are used for
identification only.

\bibliographystyle{plain}
\bibliography{references}

\end{document}